# The Impacts of Three Flamelet Burning Regimes in Nonlinear Combustion Dynamics

Tuan M. Nguyen[a,*], William A. Sirignano[a,**]

[a]*University of California, Irvine, CA 92697-3975, United States*

**Abstract**

Axisymmetric simulations of a liquid rocket engine are performed using a delayed detached-eddy-simulation (DDES) turbulence model with the Compressible Flamelet Progress Variable (CFPV) combustion model. Three different pressure instability domains are simulated: completely unstable, semi-stable, and fully stable. The different instability domains are found by varying the combustion chamber and oxidizer post length. Laminar flamelet solutions with a detailed chemical mechanism are examined. The $\beta$ Probability Density Function (PDF) for the mixture fraction and Dirac $\delta$ PDF for both the pressure and the progress variable are used. A coupling mechanism between the volumetric Heat Release Rate (HRR) and the pressure in an unstable cycle is demonstrated. Local extinction and reignition is investigated for all the instability domains using the full S-curve approach. A monotonic decrease in the amount of local extinctions and reignitions occurs when pressure oscillation amplitude becomes smaller. The flame index is used to distinguish between the premixed and non-premixed burning mode in different stability domains. An additional simulation of the unstable pressure oscillation case using only the stable flamelet burning branch of the S-curve is performed. Better agreement with experiments in terms of pressure oscillation amplitude is found when the full S-curve is used.

## 1. Introduction

In recent years, there is an increasing need for computational efficient numerical tools to simulate accurately the combustion dynamics in high-power propulsion engines such as liquid rocket engines, scramjets, and gas turbine engines. A popular method is the finite-rate chemistry model where filtered/Favre-averaged species transport equations are solved. Different approaches have been taken to address the closure problem that arises from the filtered reaction source term.

[*]Graduate Research Assistant, Corresponding author
[**]Professor

In the Laminar Closure Model (LCM), the Arrhenius reaction law is applied directly using the mean quantities [1, 2]. In the Eddy Dissipation Model [3], the reaction source terms are calculated based on turbulence quantities and different constants. In the Thickened Flame Model approach, flames are artificially thickened to be resolved on numerical grids by multiplying diffusion and dividing reaction rates by a thickening factor [4, 5]. Another approach is the Linear Eddy Mixing (LEM) model [6, 7], in which the relevant advection-diffusion-reaction couplings are resolved using a low-dimensional representation of turbulent advection. In these models, usage of any realistic detailed chemical mechanisms involving tens of species and hundreds of reactions present a difficult challenge due to the enormous computational cost. Additionally, the nonlinearity of species reaction source terms and wide range of chemical time scales associated with these schemes makes the resulting species transport equations very stiff and difficult to solve. Therefore, most of these models are limited to either one- to two-step chemical mechanisms involving 4-5 species. The transported Probability Density Function (PDF) [8, 9] is arguably the best closure models for chemistry-turbulence interaction, as it does not require any closure model for the chemical source term. However, because of the high dimensionality of its argument with Monte-Carlo simulations of at least 30-50 notional particles in a cell, the PDF simulations are usually very computationally expensive even with a simple chemistry model [8].

An alternative model to the previously mentioned models is the flamelet approach. In the flamelet concept, the chemical time scales are shorter than the turbulent time scales so that the flame can be viewed as a collection of laminar flamelets [10]. This definition allows the chemistry computation to be performed independently of the main flow simulation and pre-process as flamelet libraries/tables. Therefore, complex chemical mechanisms can be used without incurring additional computational cost on the main flow code calculations. The flamelet approach has been applied successfully to turbulent premixed flames [11, 12, 13, 14] as well as non-premixed flames [15, 16]. In the steady laminar non-premixed flamelet approach, the thermo-chemical quantities are solved in the mixture fraction space using

$$-\frac{\rho\chi}{2}\frac{\partial^2 \psi_i}{\partial Z^2} = \dot{\omega}_i \qquad (1)$$

where $\psi_i$ can be any reactive scalar quantities such as species mass fractions and temperature. The solutions of these equations can be represented by an S-curve, as shown in figure 1a.

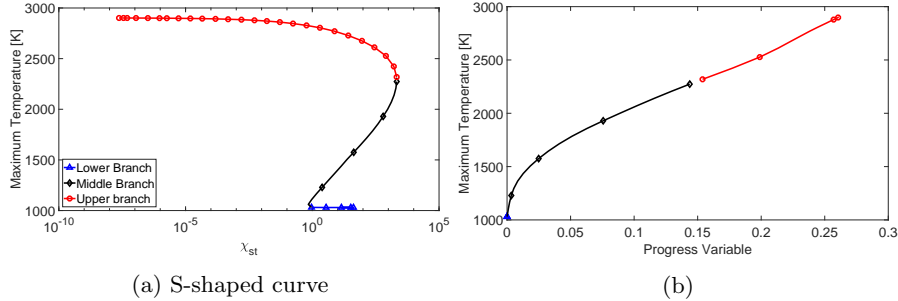

(a) S-shaped curve  (b)

Figure 1: Solutions of the steady flamelet equations for methane/oxygen combustion with $T_f = 300\ K$ and $T_o = 1030\ K$.

Figure 1a shows the maximum flame temperature as a function of the stoichiometric scalar dissipation rate $\chi_{st}$. This S-shaped curve illustrates the nature of diffusion flamelets. Each scalar dissipation rate could have multiple solutions; it is thus not a well-defined function (figure 1a). The upper branch describes stable burning solutions (curve with circle markers). The lower branch (horizontal line with triangle markers) describes non-burning solutions. The middle branch (line with diamond markers) shows the unstable burning solutions. The traditional diffusion flamelets approach of Peters [10] can only cover the upper branch. The Flamelet Progress Variable (FPV) approach, first introduced by Pierce and Moin [17], can cover all 3 branches because all the relevant quantities (e.g. maximum temperature) become monotonic functions of the progress variable (C) (figure 1b). Simulating a coaxial jet combustor, similar to the configuration used in this work, Pierce and Moin compared the FPV model to a fast-chemistry model and traditional non-premixed steady-flamelets approach. The FPV approach predicted the correct flame liftoff behavior compared with the steady flamelet approach while agreeing well with the experimental time-averaged velocities and temperature. Since the Pierce and Moin approach works, many researchers have successfully applied and extended the baseline FPV to various non-premixed and partially premixed flames. Ihme *et al.* [18, 19] studied local extinction and reignition effects in non-premixed turbulent combustion using the FPV model. The authors first compared the traditional presumed PDF ($\beta$ PDF for the mixture fraction and Dirac $\delta$ for the progress variable) with different Statistically Most Likely Distribution (SMLD) PDFs. The extended FPV model is then applied to simulations of the Sandia flames D and E. Improvements in predicting local flame extinction and reignitions compared to the baseline FPV model were found. However, priori knowledge of the SMLD PDFs is required, making it a less appealing approach compared to the baseline FPV model. Knudsen and Pitsch [20] proposed a multi-regime models by using a modified progress variable source term to distinguish between the premixed and non-premixed combustion regimes.

The works described above primarily simulate flames in the incompressible limit.

In the compressible limit, the neglect of the transient pressure effect in the flamelet formulation poses a theoretical inconsistency. However, in this work, both the time and length scales of the pressure oscillation in the chamber are much larger than those of the flamelets. Thus, a quasi-steady pressure assumption, in which the $\partial P/\partial t$ in the flamelet formulation, at any point during the pressure oscillation cycle can be justified. Moreover, the model presented below has even been applied successfully to supersonic and hypersonic combustion [21, 22, 23]. The model from here on will be called Compressible Flamelet Progress Variable (CFPV). Pecnik *et al.* [21] simulated supersonic combustions in the Hyshot II Scramjet engine using Reynolds-Averaged-Navier-Stokes (RANS) turbulence model with the CFPV combustion model. Saghafian *et al.* [22, 23] simulated combustion of a jet in a supersonic cross flow and the Hi-FiRE Scramjet engine using Large-Eddy-Simulation (LES) with the same CFPV model.

There is no combustion instability observed in any of these simulations. Additionally, to the best of the authors' knowledge, the CFPV model has not been applied to study subsonic compressible combustion. Therefore, this work examines the CPFV model capability in simulating combustion instability in a single-injector rocket engine called Continuously Variable Resonance Chamber (CVRC) [24, 25, 26, 27]. Figure 2 shows a simplified schematic of the CVRC computational domain.

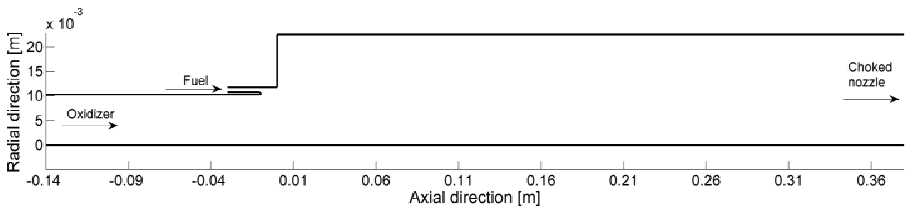

Figure 2: Computational domain for the CVRC experiments.

Different stability domains were found in the CVRC experiments by varying the oxidizer post lengths from 9 cm to 19 cm. Existing computational results using various turbulence and combustion models for these experiments are available [1, 6, 28, 29]. Srinivasan *et al.* [6] studied flame dynamics of different stability domains using LES turbulence model coupled with the LEM combustion model. Garby *et al.* [28] studied both axisymmetric and fully 3D flame stabilization mechanism for the 12-cm oxidizer post using LES method coupled with the Dynamic Flame Thickened chemistry model. Harvazinski *et al.* [1] studied the effects of grid resolution and dimensionality on the ability to predict combustion instability using both axisymmetric and 3D Detached Eddy Simulations (DES) with the LCM combustion model. Results from these simulations indicate that, while axisymmetric calculations capture the correct wave dynamics, they under-predicted the pressure oscillation amplitudes compared to three-dimensional simulations and experimental results. These simulations used either one- or two-step global chemical mechanisms. Sardeshmukh *et al.* [30] significantly improved oscillation amplitude predictions for their axisymmetric calculations by using the LCM combustion model with the GRI-Mech 1.2 detailed mechanism. However, 32 species transport equations were solved, making the computational cost prohibitively expensive.

Nguyen *et al.* [31] recently developed a computationally inexpensive axisymmetric solver utilizing the CFPV and Delay Detached Eddy Simulation (DDES) models. The code is a multi-block finite difference solver. Advection and diffusion terms are discretized using a central differencing scheme. Jameson-Schmidt-Turkel hybrid second/fourth order artificial dissipation [32] is applied for numerical stability as well as shock capturing capability. A 4-step Runge-Kutta time integration scheme is implemented. The solver is second-order accurate in space and fourth-order accurate in time. Computational cost is at least an order of magnitude lower than existing CVRC axisymmetric simulations in term of core hours per millisecond of physical time [31]. Oscillation amplitude predictions across different stability regimes agree well with experimental results. Instability mechanisms were analyzed and compared to both existing axisymmetric simulations of Garby *et al.* [28] and 3D computations of Srinivasan *et al.* [6].

Therefore, the first objective of this paper is to illustrate the importance of utilizing the whole S-curve in predicting the correct pressure oscillation amplitude. The second objective is to examine the flamelet extinction and reignition behaviors under different stability regimes. Finally, discussion regarding the combustion model ability to simulate partially premixed flame characteristic is presented. In the following sections, the numerical framework is briefly described. Readers are referred to Nguyen *et al.* [31] for complete numerical details. Flamelet solutions including small non-equilibrium effects such as oxidation and dissociation of secondary species are examined. Results and discussions of the main CFD computations are followed. Finally, conclusions are presented.

## 2. Numerical Framework

### 2.1. Governing equations

For a multispecies mixture, the Favre-Averaged Navier-Stokes equations are written in conservative form following [21]

$$\frac{\partial \bar{\rho}}{\partial t} + \frac{\partial \bar{\rho}\tilde{v}_j}{\partial x_j} = 0 \tag{2}$$

$$\frac{\partial \bar{\rho}\tilde{v}_i}{\partial t} + \frac{\partial \bar{\rho}\tilde{v}_i\tilde{v}_j}{\partial x_j} = -\frac{\partial \bar{p}}{\partial x_i} + \frac{\partial(\tau_{ij} + \tau_{ij}^R)}{\partial x_j} \tag{3}$$

$$\frac{\partial \bar{\rho}\tilde{E}}{\partial t}+\frac{\partial \tilde{v}_j(\bar{\rho}\tilde{E}+\bar{p})}{\partial x_j} = \frac{\partial}{\partial x_j}\left[\tilde{v}_i(\tau_{ij}+\tau_{ij}^R)+\left(\mu+\sigma_k\mu_t\frac{\partial k}{\partial x_j}\right)\right]+\frac{\partial}{\partial x_j}\left[\left(\frac{\lambda}{c_p}+\frac{\mu_t}{Pr_t}\right)\frac{\partial \tilde{h}}{\partial x_j}\right] \quad (4)$$

where $\bar{\rho}$ is the mean density, $u_i$ is the velocity in the $x_i$ direction. $\bar{p}$ is the mean pressure. $\mu$ and $\mu_t$ are the molecular and turbulent viscosity. $\lambda$ and $C_p$ are the heat conduction and constant specific heat coefficients. $\tau_{ij}$, $\tau_{ij}^R$ are the molecular and turbulent viscous stress tensors, respectively:

$$\tau_{ij} = \mu\left(\frac{\partial \tilde{v}_i}{\partial x_j}+\frac{\partial \tilde{v}_j}{\partial x_i}-\frac{2}{3}\frac{\partial \tilde{v}_k}{\partial x_k}\delta_{ij}\right) \qquad \tau_{ij}^R = \mu_t\left(\frac{\partial \tilde{v}_i}{\partial x_j}+\frac{\partial \tilde{v}_j}{\partial x_i}-\frac{2}{3}\frac{\partial \tilde{v}_k}{\partial x_k}\delta_{ij}\right) \quad (5)$$

The total energy, $\tilde{E}$, has the form of

$$\tilde{E} = \frac{1}{2}\left(\sum_{j=1}^{n}\tilde{v}_j\tilde{v}_j\right)+k+\tilde{e} \quad (6)$$

where n is the number of dimensions. The first term on the right side is the mean flow kinetic energy. The second term is the turbulent kinetic energy, $k$. $\tilde{e}$ is the total thermal energy which includes the sensible and chemical energies. Enthalpy, $\tilde{h}$, is related to the total thermal energy as $\tilde{h} = \tilde{e} + \frac{\bar{p}}{\bar{\rho}}$. For high pressure combustion, the ideal gas law is assumed ($\bar{p} = \bar{\rho}R\tilde{T}$), where R is the specific gas constant. The turbulent Schmidt ($Sc_t$) and Prandtl ($Pr_t$) numbers are assumed to be constant at 0.9 [21].

2.2. Turbulence model

Here, the DDES model is based on the 2006 Wilcox $k-\omega$ model [33]. The conservative form of the governing equations for the turbulent kinetic energy ($k$) and the turbulent disspation rate ($\omega$) are written as follows [33]

$$\frac{\partial \bar{\rho}k}{\partial t}+\frac{\partial(\bar{\rho}\tilde{v}_j k)}{\partial x_j} = (\tau_{ij}+\tau_{ij}^R)\frac{\partial \tilde{v}_i}{\partial x_j}-\beta^*\bar{\rho}\omega k+\frac{\partial}{\partial x_j}\left[\left(\mu+\sigma_k\frac{\rho k}{\omega}\right)\frac{\partial k}{\partial x_j}\right] \quad (7)$$

$$\frac{\partial \bar{\rho}\omega}{\partial t}+\frac{\partial(\bar{\rho}\tilde{v}_j\omega)}{\partial x_j} = \frac{\gamma\omega}{k}(\tau_{ij}+\tau_{ij}^R)\frac{\partial \tilde{v}_i}{\partial x_j}-\beta\bar{\rho}\omega^2+\frac{\partial}{\partial x_j}\left[\left(\mu+\sigma_\omega\frac{\bar{\rho}k}{\omega}\right)\frac{\partial \omega}{\partial x_j}\right]+\frac{\bar{\rho}\sigma_d}{\omega}\frac{\partial k}{\partial x_j}\frac{\partial \omega}{\partial x_j} \quad (8)$$

where $\beta^*, \beta, \sigma_k, \sigma_\omega, \sigma_d$ are modeling constants. The turbulent viscosity is calculated as $\mu_t = \frac{\bar{\rho}k}{\hat{\omega}}$, where $\hat{\omega}$ is the turbulent frequency corrected by the maximum of $\omega$ and the flow mean strain rate.

In any version of the Detached Eddy Simulation (DES) approach, the dissipation term in equation 7 (second term on the right side) is modified to exclude any grid-realized contribution in the turbulent viscosity. This can be achieved using the mathematical definition [34]:

$$\beta^* \rho \omega k = \rho k^{3/2}/L_T^* \qquad (9)$$

where the corrected turbulent length scale is defined as

$$L_T^* = min(L_T, C_{DES} L_{GRID}) \qquad (10)$$

where $C_{DES}$ is a modeling constant, and $L_T$ and $L_{GRID}$ are the turbulent and grid length scales where $L_T = k^{1/2}/(\beta^* \omega)$ and $L_{GRID} = L_T - F_D(L_T - \Delta)$. $\Delta$ is the largest grid dimension for the cell. In the traditional DES approach, $F_D$ has a value of unity. In the DDES approach, $F_D$ is a hyperbolic tangent blending function which uses the distance of the cell away from the wall as an input [35]. This blending function is used to limit grid arbitrariness because the smallest grid sizes of the mixing shear layer and walls are the same in this work.

2.3. Compressible Flamelet Progress Variable Approach

In the CFPV approach [23], presumed PDFs are used to relate the laminar flamelet solutions in the mixture fraction space to their Favre-averaged/mean counterparts. The $\beta$ PDF is assumed for the mixture fraction while the Dirac $\delta$ PDF is assumed for both the progress variable and pressure. The Favre-averaged thermo-chemical quantities ($\tilde{\psi}_i$) at each pressure value are pre-processed as lookup libraries using the convolution:

$$\tilde{\psi}_i(\tilde{Z}, \widetilde{Z''^2}, \tilde{C}, \bar{p}) = \int_0^1 \int_0^C \int_{p_o}^p \psi_i(Z, C) \beta(Z, Z''^2) \delta(C) \delta(p) dZ dC dp \qquad (11)$$

where Z is the mixture fraction, C is the progress variable. In this work, the progress variable is defined as the total mass fraction of $H_2$ and $CO_2$. During the CFD computation, the transport equations for the mean scalars $\tilde{Z}, \widetilde{Z^2}, \tilde{C}$ are solved. The mean mixture fraction squared ($\widetilde{Z^2}$) is related to the mean mixture fraction and the mean variance as: $\widetilde{Z^2} = \tilde{Z}^2 + \widetilde{Z''^2}$. With Lewis number equal to one, the transport equations for these scalars are given as

$$\frac{\partial \bar{\rho} \tilde{Z}}{\partial t} + \frac{\partial \bar{\rho} \tilde{v}_j \tilde{Z}}{\partial x_j} = \frac{\partial}{\partial x_j}\left[\left(\frac{\lambda}{c_p} + \frac{\mu_t}{Sc_t}\right)\frac{\partial \tilde{Z}}{\partial x_j}\right] \qquad (12)$$

$$\frac{\partial \bar{\rho} \widetilde{Z^2}}{\partial t} + \frac{\partial \bar{\rho} \tilde{v}_j \widetilde{Z^2}}{\partial x_j} = \frac{\partial}{\partial x_j}\left[\left(\frac{\lambda}{c_p} + \frac{\mu_t}{Sc_t}\right)\frac{\partial \widetilde{Z^2}}{\partial x_j}\right] - \bar{\rho} C_x \omega (\widetilde{Z^2} - \tilde{Z}^2) \qquad (13)$$

$$\frac{\partial \bar{\rho}\tilde{C}}{\partial t} + \frac{\partial \bar{\rho}\tilde{v}_j\tilde{C}}{\partial x_j} = \frac{\partial}{\partial x_j}\left[\left(\frac{\lambda}{c_p} + \frac{\mu_t}{Sc_t}\right)\frac{\partial \tilde{C}}{\partial x_j}\right] + \tilde{\bar{\omega}}_C \qquad (14)$$

where $C_x$ has a constant value of 2.0 [21]. Turbulence mixing and turbulence/flame interaction of the mean mixture fraction are modeled by solving equation (13), which implicitly describes the variance of the mean mixture fraction ($\widetilde{Z''}$) [36].

At each time step, the local values of these scalars along with the pressure allow us to retrieve quickly properties such as local compositions, temperature, specific heat ($c_p$), enthalpy, and thermal diffusivity using pre-tabulated flamelet libraries.

At the end of each time step, the local values of the Favre-averaged thermal energy ($\tilde{e}$) can be different from the thermal energy ($e_f$) computed from the turbulent flamelet transport equations (equations 12-14). However, the local compositions are the same for both quantities. Following Pecnik *et al.* [21], for a given $\tilde{e}$ value computed from the Navier-Stokes equations, an expansion around the thermal energy of the flamelet solutions has the form

$$\tilde{e} = e_f + \int_{T_f}^{\tilde{T}} c_v(T)dT = e_f + \int_{T_f}^{\tilde{T}} \frac{R_f}{\gamma(T)-1} dT \qquad (15)$$

where the subscript "f" denotes the values of the flamelet solution. The specific heat ratio ($\gamma$) can be expressed as:

$$\gamma(\tilde{T}) = \gamma_f + a_\gamma(\tilde{T} - T_f) \qquad (16)$$

where $a_\gamma$ is the local linear expansion coefficient and tabulated during the pre-processing step as a flamelet library.

Integrating equation (15) and solving for $\tilde{T}$ we get

$$\tilde{T} = T_f + \frac{\gamma_f - 1}{a_\gamma}\bigg(\exp(a_\gamma(\tilde{e} - e_f)/R_f) - 1\bigg) \qquad (17)$$

Equation (17), together with the ideal gas law, illustrate the nonlinear coupling between the flame and the acoustical field.

*2.4. CVRC Details*

As seen in figure 2, the CVRC is essentially a coaxial dump combustor. The oxidizer is injected in the central tube and fuel is injected in the concentric outer tube. The fuel is methane with a temperature of 300 K. The oxidizer is decomposed hydrogen peroxide with a composition of 58% $H_2O$ and 42% $O_2$ by mass. The oxidizer temperature is 1030 K. In all cases considered in this paper, fuel and oxidizer mass flow rates are held constant at 0.027 kg/s and 0.32 kg/s, respectively. The mass oxidizer-to-fuel ratio based on the inlet flow rates is 11.85. With a stoichiometric oxidizer-to-fuel ratio of 9.52, the flow is globally fuel lean with an equivalence ratio of 0.8.

All walls are adiabatic, impermeable, and no-slip. The constant mass flow rate inlet boundary condition is implemented using the Navier-Stokes Characteristic Boundary Conditions [5]. To save computational resources, a short-choked-nozzle [37] outlet boundary condition is used instead of an actual convergent-divergent nozzle computational domain. Based on the CVRC experimental geometry, the entrance-to-throat area ratio is 5. Compared to the results shown in Nguyen *et al.* [31], more grid points are placed in the mixing shear layer and around the dump plane to capture better the thin reaction zones region. The resulting mesh consists of more than 139000 grid points across all cases.

### 3. Results and discussions

#### 3.1. Flamelet solutions

The laminar flamelet solutions are solved using the FlameMaster code [38]. A 72-reaction detailed mechanism with 27 species (neglecting nitrogen) is used [39]. Figure 3a shows different temperature solutions along the S-curve. When the mixture fraction is zero, the flow is solely composed of oxidizer. When the mixture fraction is one, the flow is solely composed of fuel. Therefore, the left boundary of all the temperature curves is always 1030 K and their right boundary is 300 K. The bottom curve represents mixing branch. The next 6 curves above it represent unstable flamelet burning solutions, along the middle branch of the S-curve. These solutions are classified as unstable due to their sensitivity to small perturbations by moving either toward the stable upper branch or toward a stable quenched solution [20]. The top 3 curves represent stable flamelet burning solutions. Figure 3b shows the maximum HRR as functions of the progress variable at different pressures. The total HRR rate for n number of species is defined as

$$HRR = -\sum_{k=1}^{n} \dot{\omega}_k h_k \qquad (18)$$

where $\dot{\omega}_k$ is the mass reaction rate per unit volume ($kg/m^3s$) of the k species. $h_k$ is the species enthalpy ($J/kg$), which also includes the enthalpy of formation). Depending on the pressure, the progress variable values at extinction is from 0.145-0.155. Pressure effects on the flame can clearly be observed from figure 3b. Since the Dirac delta function is the marginal PDF for pressure, the mean pressure ($\bar{p}$) is the same as the background pressure in the flamelet solutions. Using this fact along with figure 3b, the coupling relationship between the pressure and HRR can now be clearly observed. As the pressure increases, the HRR increases as there is more mass per unit volume to burn. Admittedly the effect of using the Dirac delta function for the pressure remains somewhat ambiguous. However, by using equation 17 to obtain the mean temperature and solving for the full Navier-Stokes equations, we allow the pressure waves to propagate independently (to a certain extent) from the flamelet model.

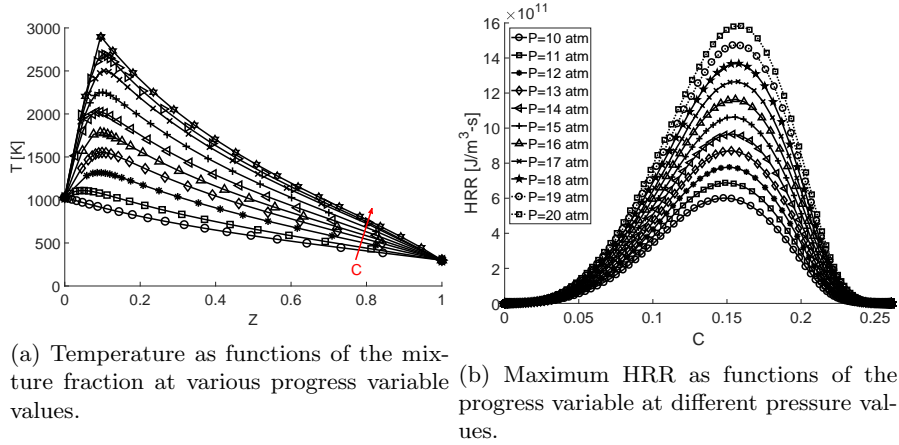

(a) Temperature as functions of the mixture fraction at various progress variable values.

(b) Maximum HRR as functions of the progress variable at different pressure values.

Figure 3: Representative flamelet solutions for the entire S-curve.

Figure 4 shows two different flamelet solutions for approximately the same reference stoichiometric dissipation rate near the quenching limit. The left column represents an unstable flamelet burning solution while the right column represents a stable flamelet burning solution. The top row shows the HRR as well as the HRR by major species. The bottom row shows major species mass fractions as well as temperature profile. The region enclosed by the vertical lines is the approximated oxidation layer [40]. The right vertical lines mark the stoichiometric mixture fraction.

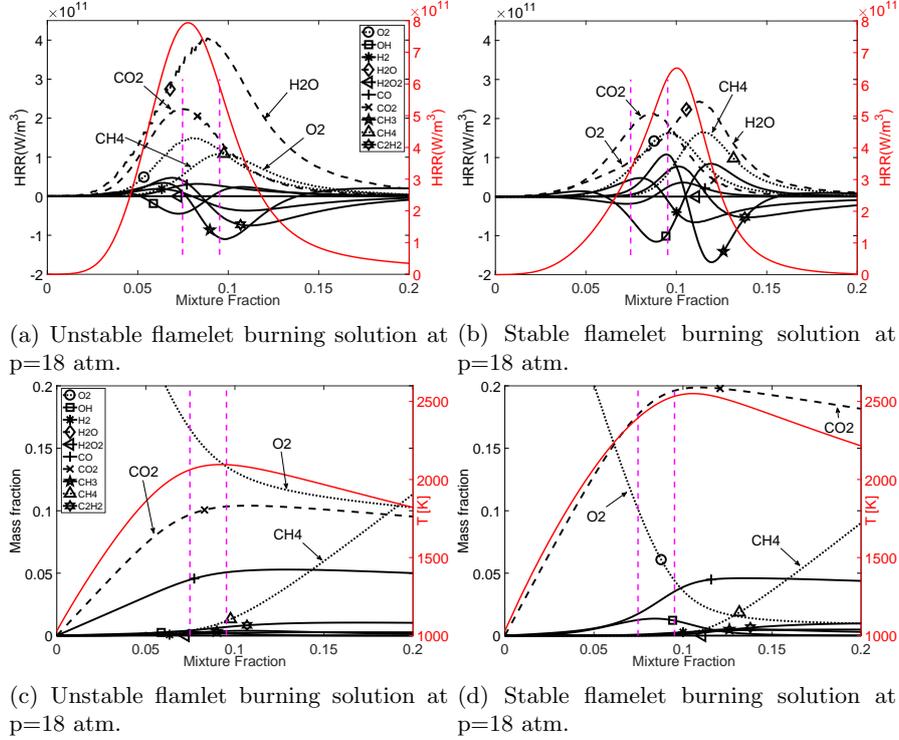

Figure 4: Major species reaction rates, temperature, and HRR as functions of the mixture fraction.

In the unstable flamelet burning limit, HRR on the fuel-lean side is dominated by the non-equilibrium effects, as seen from figure 4a. The most intense HRR region thus locates within the oxidation layer. In the stable flamelet burning limit, however, the HRR region is located on the fuel-rich side, slightly to the right of the stoichiometric line. Non-equilibrium structure within the oxidation layer no longer has a significant effect. In both cases, due to high scalar dissipation rates in the reaction zone (small characteristic diffusion time), there is significant reactant leakage through the reaction zone, as shown in figures 4c-4d. In the stable burning case, oxygen is consumed faster across the oxidation layer, thus leading to less oxygen leakage to the fuel-rich side. The flame structures described above are similar to findings by Seshadri and Peters [40] for methane-air diffusion flame. The above analysis is meant to demonstrate the effectiveness of the current flamelet formulation in capturing the correct flame structures. The readers are referred to Jorda-Juanos and Sirignano [41] and Wang *et al.* [42] for complete descriptions of methane/oxygen diffusion flames at high pressure.

*3.2. Combustion Instability in the CVRC*

In the CVRC experiments, a different instability regime is found by varying the oxidizer post length. In the following sections, three different combustion chamber instability regimes are described: fully unstable with a limit cycle, semi-stable, and fully stable. The fully unstable limit-cycle behavior occurs with the 14-cm oxidizer post and 38-cm chamber configuration. The peak-to-peak pressure oscillation amplitude is 600 kPa. The semi-stable behavior occurs with the 9-cm oxidizer post and 38-cm chamber configuration. The peak-to-peak pressure oscillation amplitude in this case is 200 kPa. The stable behavior is found with the 17-cm oxidizer post and 30-cm chamber configuration. The peak-to-peak oscillation amplitude is around 80 kPa, which is about 5% of mean chamber pressure (P=1540 kPa). This case is therefore classified as stable. Using Power Spectral Density (PSD) analysis, the first-mode frequencies of the 9-cm, 14-cm, and 17-cm cases are 1400 Hz, 1520 Hz, and 1622 Hz. With a mean chamber value of approximately 1700 kPa, the 14-cm oxidizer post case has the widest operating pressure from 1400-2100 kPa, which is well under the critical pressure values for most reactants except $H_2$ (table 1). Huo and Yang [43] show that, for supercritical combustion of oxygen/hydrogen mixtures, the ideal gas law assumption has negligible effect on the flame structure. Therefore, the assumption of ideal gas law is valid across all cases.

Table 1: Critical properties of different reactants

| **Reactants** | $T_{cr}$ (K) | $P_{cr}$ (atm) |
|:---:|:---:|:---:|
| $CH_4$ | 190.6 | 45.6 |
| $O_2$ | 154.6 | 49.8 |
| $H_2$ | 33.2 | 12.8 |
| $H_2O$ | 647 | 217.75 |
| $CO_2$ | 304.18 | 72.83 |
| $CO$ | 134.45 | 34.98 |

Figure 5 shows the oscillatory behaviors and first-longitudinal-mode shape for all cases. The mode shape is obtained by computing the modulus of the unsteady pressure signals along the longitudinal axis based on the first-mode frequency identified in the PSD analyses. Half-wavelength standing waves occur in the chamber across all cases. Pressure nodes are found approximately in the middle of the chamber. Clear limit-cycle behavior is observed for the 14-cm case.

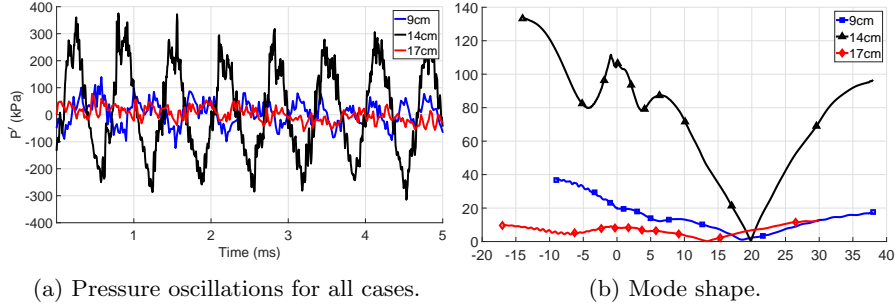

(a) Pressure oscillations for all cases.  (b) Mode shape.

Figure 5: Pressure oscillation behavior and first longitudinal-mode-shape for the 14-cm oxidizer post case.

An important measurement of combustion instability is the Rayleigh index, which is a correlation used to determine the locations where the pressure oscillations are driven or damped by the unsteady HRR. The time-averaged spatially local Rayleigh index [28] is defined as

$$RI = \frac{1}{\tau}\int_{t_o}^{t_o+\tau} \frac{\gamma-1}{\gamma} p'\dot{\omega}' dt \qquad (19)$$

where $p'$ is the local pressure oscillation and $\dot{\omega}'$ is the local HRR oscillation. Positive Rayleigh index indicates the pressure oscillations are driven by the unsteady HRR. Figure 6 presents the Rayleigh index result for the 14-cm case. Only half of the combustion chamber is shown.

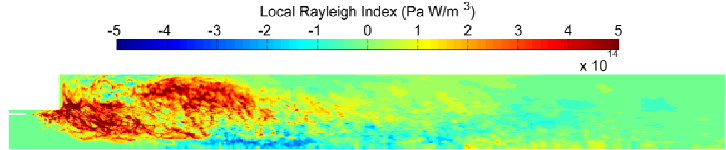

Figure 6: Time-averaged spatially local Rayleigh index.

Figure 6 shows a strong correlation between the pressure oscillations and the HRR around the recirculation zone as well as the mixing layer immediately after the splitter plate. Along with figure 5, the main region for the instability driving mechanism in the CVRC experimental rig is now clearly observed. Evidently, the strong coupling between the HRR and pressure oscillations is supported by the coupling location around the pressure anti-node (back-step), thus further promoting the instability. In the 9-cm case, the maximum positive Rayleigh index values are an order of magnitude smaller than shown here for the 14-cm case, and only found around the mixing shear-layer immediately downstream of the back step. The readers are referred to Nguyen *et al.* [31] for the Rayleigh index results of the 9-cm case. The 17-cm case Rayleigh-index result indicates weak pressure-HRR coupling along the chamber. It is, however, not plotted here for brevity purposes. The Rayleigh-index analysis is further supported by the pulse-timing mechanism described in Nguyen *et al.* [31].

*3.3. Flame Dynamics*

Figure 7 illustrates the local ignition and extinction effects for the 14-cm case. The white isocontour lines represent the stoichiometric mixture fraction value ($\tilde{Z}_{st} = 0.095$). Only the region between the splitter plate and the combustor dump plane is shown here. The spatial vectors are plotted in millimeter. Four random probes are placed in each figure. These probes are now numbered I to IV following from left to right and counterclockwise direction. Therefore, probe I is located at ($x = -6.6\ mm$, $r = 8.9\ mm$) and probe IV is located at ($x = -2.6\ mm$, $r = 10.2\ mm$). Local extinctions and reignitions are clearly observed by following the transient behavior probe I. At time t1 (first row of figure 7), probe I is fuel-rich ($\tilde{Z} = 0.27$). Moderately burning ($HRR = 164\ GJ/m^3s$) occurs in the unstable flamelet burning branch ($\tilde{C} = 0.14 < 0.16$). At time t2 (second row), the location is fully burning on the stable branch. The HRR increases by a factor of three while the fuel-rich mixture is still maintained. At time t3 (the third row), while the local flow composition remains relatively the same compared to time t2, the flame is locally extinct with its HRR decreased by a factor of 10. Further examination shows an increase in the Favre-averaged scalar dissipation rate ($\tilde{\chi}$) at this probe from time t2 to t3 (from 4251 $1/s$ to 7776 $1/s$ ). Thus the flame is extinguished. For similar values of $\tilde{Z}$ and C, the increase in the mean scalar dissipation rate is caused by the decrease in the mean mixture variance, mainly through the transport of the $\widetilde{Z^2}$ equation. At time t5 and still following probe I, the flame is now locally ignited, but burning in the unstable burning branch ($\tilde{C} = 0.10 < 0.16$). The $\tilde{Z}$ value significantly decreases from time t4 to t5 due to the propagation of the unburnt reactant mixtures.

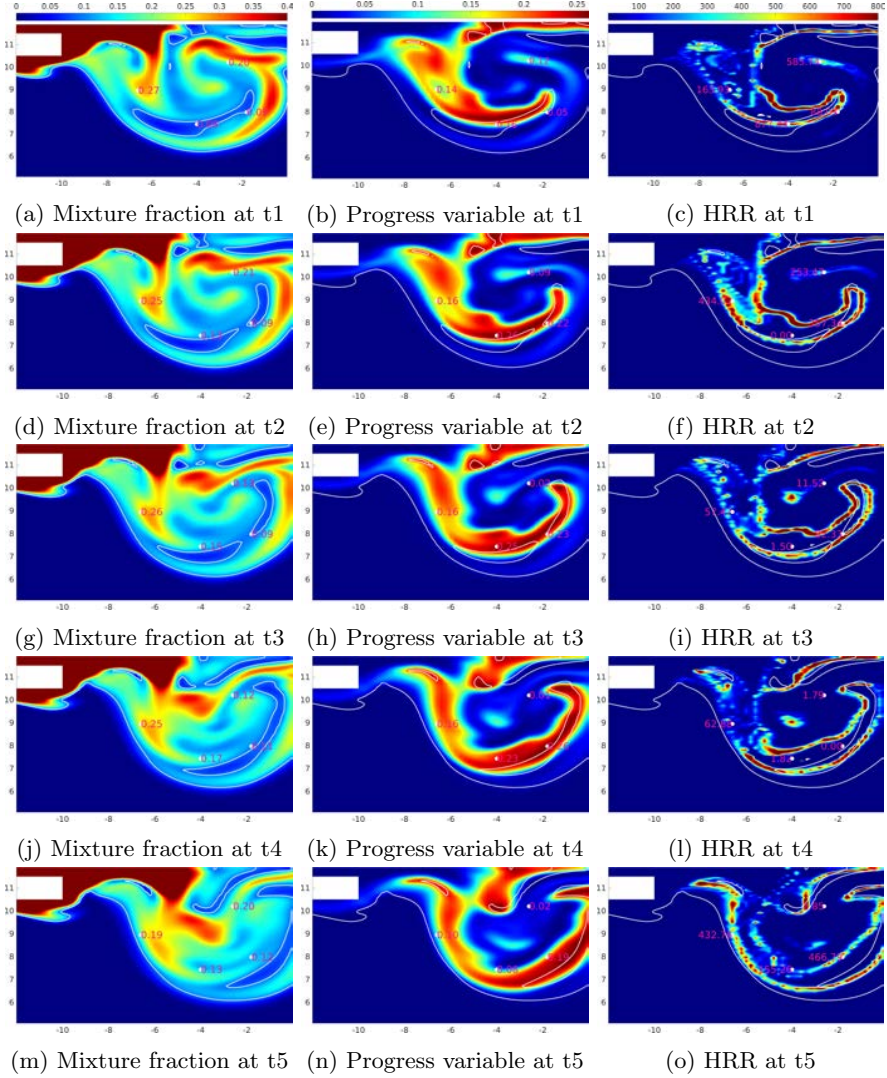

Figure 7: Unsteady behaviors of different flow variables subjected to an adverse pressure gradients during the peak of the unstable pressure oscillation. Each time frame is separated by 5 microseconds. HRR has a unit of $GJ/m^3s$.

Following probe II, at time t1, there is strong non-equilibrium burning, in which the flame is substantially fuel-lean ($\tilde{Z} = 0.08$) even though it is burning in the stable flamelet burning branch. Due to its non-equilibrium structure, the flame burns in a very short time ($\tilde{C} = 0.16$ to $\tilde{C} = 0.26$ ). A small flame burning on the fuel-rich side in the unstable flamelet burning branch is found at this location at time t5. At time t1 for probe IV, a small, intense flame (hot spot) is found burning in the unstable flamelet burning branch. This hot spot continuously burns for at least 15 microseconds as it moves to the center of the roll-up vortex.

Examination beyond time t5 reveals that once the left-running pressure wave passes toward the oxidizer inlet upstream of the splitter plate, the roll-up vortex move significantly faster downstream toward the combustion chamber. The flame in the region shown in figure 7 burns weakly and is mostly diffusion controlled, as shown in figure 8 which is 50 microseconds later than time t5. The readers are referred to Nguyen *et al.* [31] for a complete limit cycle behavior analysis.

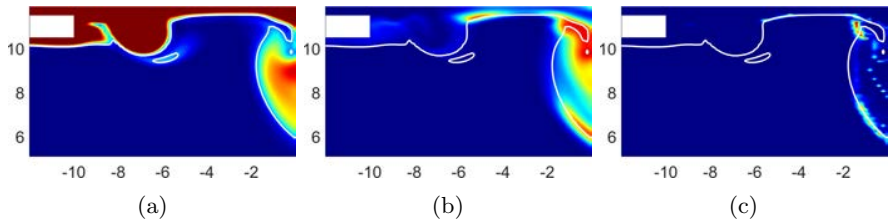

(a)  (b)  (c)

Figure 8: Contour plots of the mean mixture fractions (left column), the progress variable (middle column), and the HRR (right column) subjected to a favorable pressure gradient during the trough of the unstable pressure oscillation. The same scales are used when compared to figure 7.

To further illustrate the importance of utilizing the whole S-curve in the calculation, an additional simulation of the 14-cm case is performed. While the full flamelet equations are solved, the progress variable values are limited to above 0.16. The flame in this case, therefore, can only burn in the stable flamelet burning branch. This approach is similar to the Steady Laminar Flamelet Model [10]. The peak-to-peak pressure oscillation amplitude is 200-250 kPa, which is roughly one-third of the amplitude predicted when the full S-curve is allowed. In comparison, the experimental peak-to-peak amplitude is 750 kPa for this case. This pressure oscillation amplitude matches well with calculations in Pant *et al.* [44] using the Steady Laminar Flamelet Model. The flame in this case is much cooler compared to the full S-curve simulation, as seen in figure 9c. The flame is strongly attached to the splitter plate regardless of whether favorable or adverse pressure gradients are imposed on the flow. Similar phenomena are observed for a coaxial combustor simulation using the steady laminar flamelet approach [17]. The lack of local extinctions and reignitions also means the flame front cannot be lifted and reattached as seen when the full S-curve is allowed. The Rayleigh index analysis in this case (figure 9d) reveals significant reduction in pressure-HRR coupling around the dump plane (pressure anti-node).

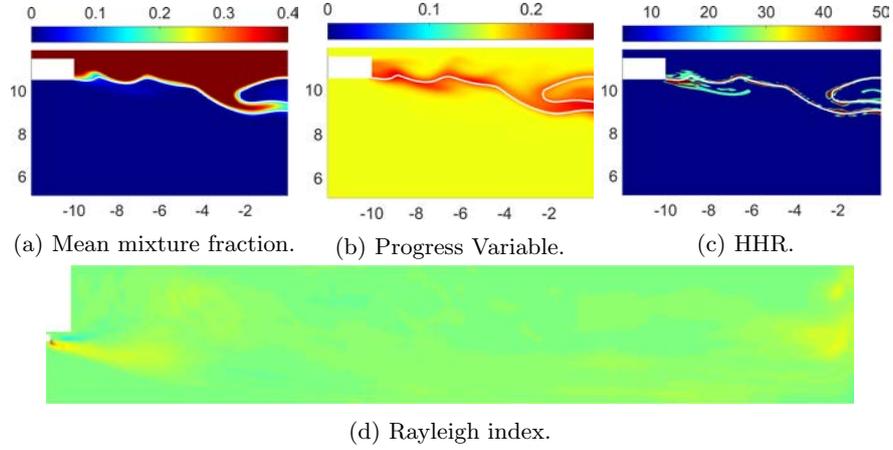

(a) Mean mixture fraction.    (b) Progress Variable.    (c) HHR.

(d) Rayleigh index.

Figure 9: Contour plots of the simulation results using only the stable flamelet branch. HRR has a unit of $GJ/m^3 s$. The Rayleigh index is plotted using the same scale as in figure 6.

Figure 10 shows the HRR and fuel consumption rate for each of three cases. The flame occurs much closer to the injector lips (upstream of the dump plane) compared with the stable and semi-stable cases. The flame in the stable case is lifted further away from the dump plane compared to the semi-stable case. The most intense fuel consumption region does not completely overlap the high HRR region. Specifically, in the recirculation zone of case 14-cm, there are regions in which fuel consumption rate is low but high HRR are found. These regions are dominated by the non-equilibrium flame structures. This phenomenon allows stronger coupling between the HRR and the pressure, leading to higher pressure amplitudes prediction (and closer to the experimental results) compared to other existing axisymmetric calculations.

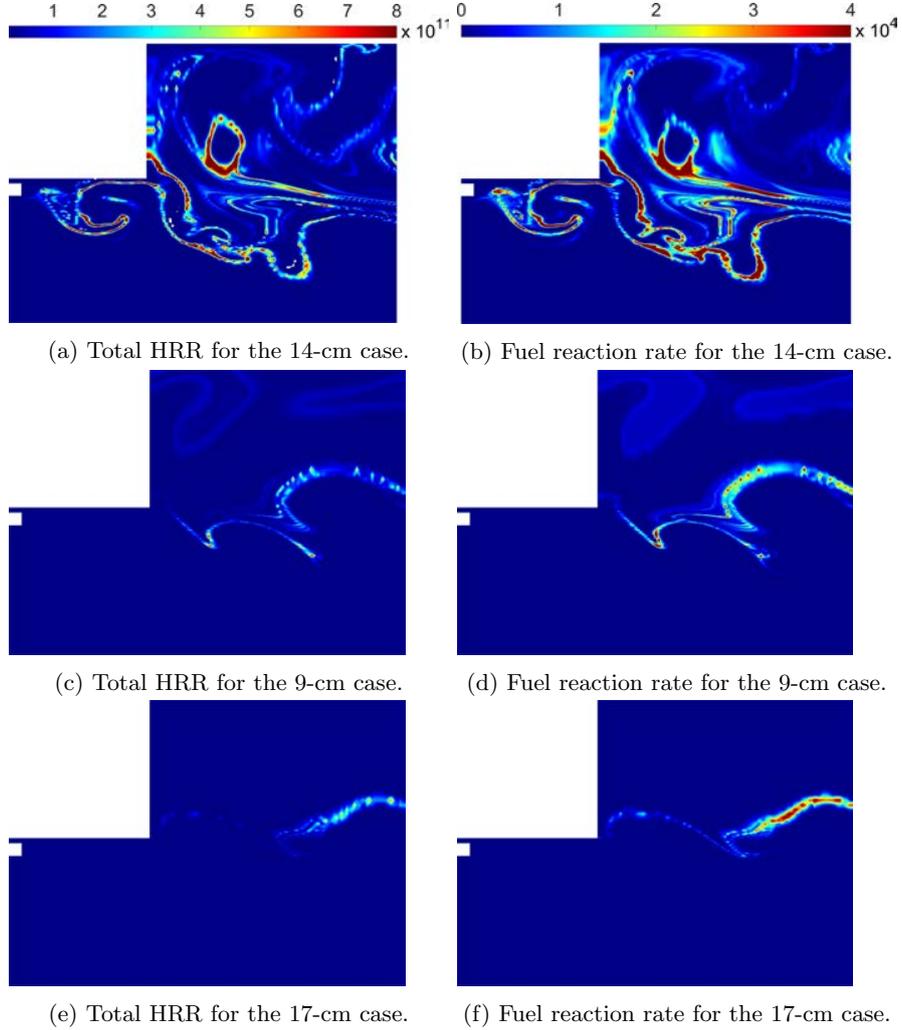

(a) Total HRR for the 14-cm case.   (b) Fuel reaction rate for the 14-cm case.

(c) Total HRR for the 9-cm case.   (d) Fuel reaction rate for the 9-cm case.

(e) Total HRR for the 17-cm case.   (f) Fuel reaction rate for the 17-cm case.

Figure 10: HRR ($J/m^3 s$) and fuel consumption rate ($kg/m^3 s$) contour plots for three different cases.

The nonlinearity of the pulsing mechanism as well as high axial-to-radial aspect ratio presents a difficult challenge in the examination of the transient behaviors of local extinction and ignitions such as the one shown in figure 7. Scatter plots of the mean temperature somewhat alleviate the difficulty. Figure 11 shows scatter plots of the Favre-averaged temperatures ($\tilde{T}$) as functions of the mean mixture fractions ($\tilde{Z}$) at two different locations of the combustion chamber.

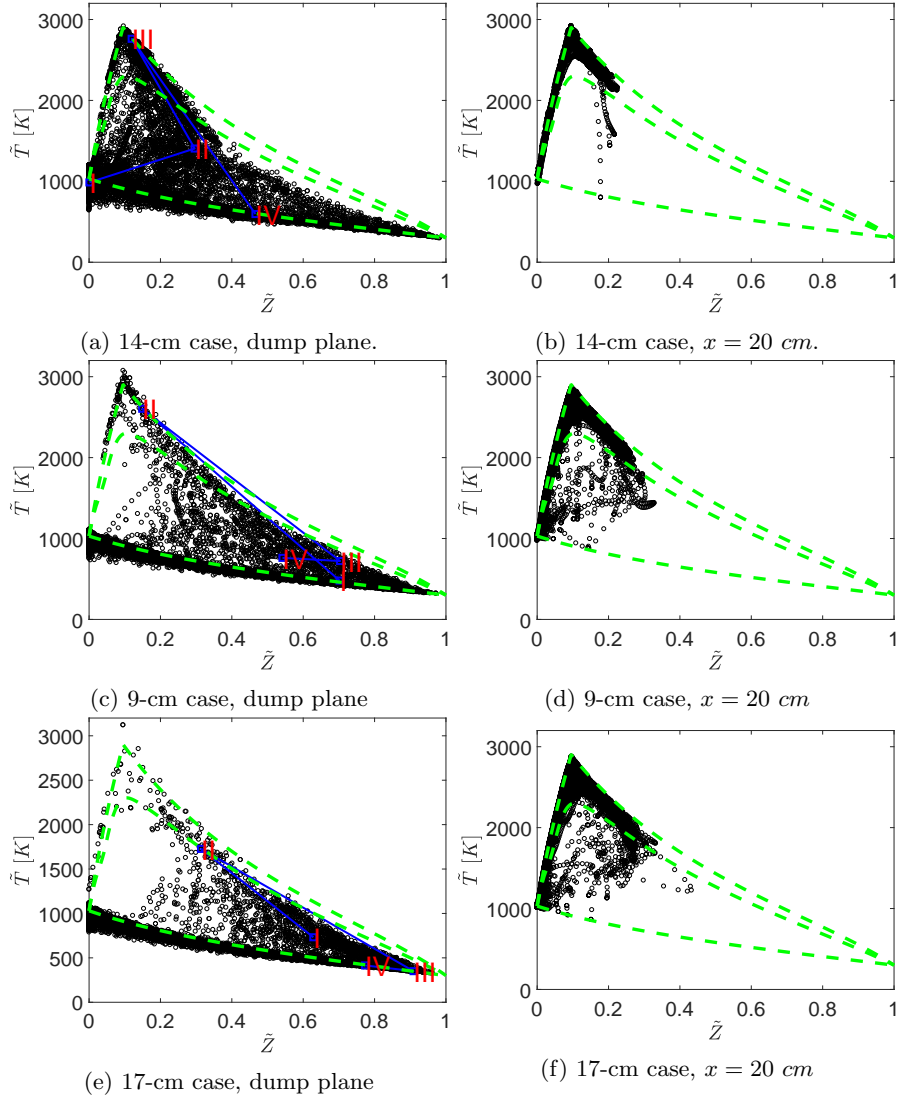

Figure 11: Scatter plots of the Favre-averaged temperature as functions of mean mixture fraction. Solutions from the laminar flamelet library are also shown as the broken lines.

The broken lines in these figures represent the laminar-flamelet solutions along the S-curve, similar to figure 3a. The middle broken lines represent the flame solution at the quenching limit between the stable and unstable branches on the S-curve. The first column represents the data sampled over multiple oscillation periods at the dump plane of the combustion chamber ($x = 0\ cm$). The second column represents the data of the entire chamber surface (a line in our axisym-metric calculations) at the midpoint of the combustion chamber ($x = 20\ cm$). This location represents the pressure node for both the 14- and 9-cm case. Examining the reaction zone of the first column in figure 11, it is clear that more burning occurs much closer to the injector lips as the oscillation amplitude increases. Moreover, local extinctions and reignitions occur much more frequently and intensely in the reaction zone of the 14-cm case compared to the other two cases. This point is further illustrated by the blue solid line in these figures. These lines represent the transient behaviors of an arbitrary point within the shear layer at the dump plane. The points along these curves are labeled by Roman numerals, indicating their orders in time. As seen from figure 11a, at time t= *I*, the first point represents an unmixed oxygen situation. At time t= *II*, an unstable flame is observed. The flame is fully burning at time t= *III*. It is then extinguished at time t= *IV*, thus returning it to the mixing line. Similar behavior can be observed for the 9-cm case but on the fuel-rich side. On the other hand, in the 17-cm case, the flame could only burn in the unstable burning branch. The flame is thus strongly anchored at the back step for the 9-cm case while completely lifted from the back step in the 17-cm case. Comparing the second column of figure 11, the flame is fully burning on the stable branch in the 14-cm case. In the other two cases, the flame still burns in both stable and unstable branches, but with much less intense local ignitions and extinctions compared to the dump plane of the 14-cm case. Finally, by comparing the first row of figure 11, we can see the influence of the pressure oscillations on the flamelet temperatures. Particularly, at $\tilde{Z} = 0$, at the dump plane (a pressure anti-node), the temperature of the oxidizer stream can differ by more than 300 K from the laminar flamelet solution. On the other hand, at $x = 20\ cm$, the oxidizer temperature remains close to the flamelet solution.

## 4. Burning mode

As shown previously, the flame in the CVRC is classified as partially premixed regardless of its instability characteristics [31, 6]. Previously, Nguyen *et al.* [31] used the following flame index definition to distinguish between the premixed and non-premixed burning mode:

$$FI = \frac{\nabla Y_f \cdot \nabla Y_o}{\left|\nabla Y_f\right|\left|\nabla Y_o\right|}|\dot{\omega}_f| \qquad (20)$$

where $Y_o$, $Y_f$ are the oxidizer and fuel mass fraction, respectively. $\dot{\omega}_f$ is the fuel consumption rate. In equation (20), the first term is the classical Takeno flame index. Therefore, the flame index is positive (premixed burning) when the reactant gradients are aligned and negative (non-premixed burning) when the react gradients are opposite of each other. It was shown in figure 10, however, there exists high HRR region due to nonequilibrium effects on the fuel-lean side even when the fuel consumption rate is small. Therefore, the flame index will be modified as

$$FI = \frac{\nabla Y_f \cdot \nabla Y_o}{\left|\nabla Y_f\right|\left|\nabla Y_o\right|} \dot{\omega}_T \qquad (21)$$

where $\dot{\omega}_T$ is the HRR. In the following analysis, the flame index is first computed, then volume-averaged over the combustion chamber (including the upstream splitter plate region). Figure 12 shows the total volume-averaged HRR of the combustion chamber and its fraction that is burning in a non-premixed mode (taking only the negative flame index value.). For clarity, only 2.1 milliseconds are shown in each figure, which corresponds to roughly three first-mode pressure oscillation cycle in the 14-cm case. The initial time in the 14-cm case corresponds to time t1 in figure 7. The dominant burning mode is now premixed, as shown by the flame dynamics analysis. As the pressure in the chamber drops, the flame moves further downstream while become non-premixed dominant. This observation is further supported by figure 8. The cycle repeated itself roughly every 0.7 millisecond, correspond to a first-mode cycle period with a frequency of 1520 Hz for the 14-cm case. In the 9-cm case, the lack of a strong pressure oscillation leads to less fluctuation in the averaged HRR. The 17-cm case exhibits similar behavior to the 9-cm case but with less fluctuation in its burning mode because of its stable pressure behavior. The averaged fractions over time of non-premixed burning mode are 46%, 41%, 38% for the 14-cm, 9-cm, and 17-cm cases.

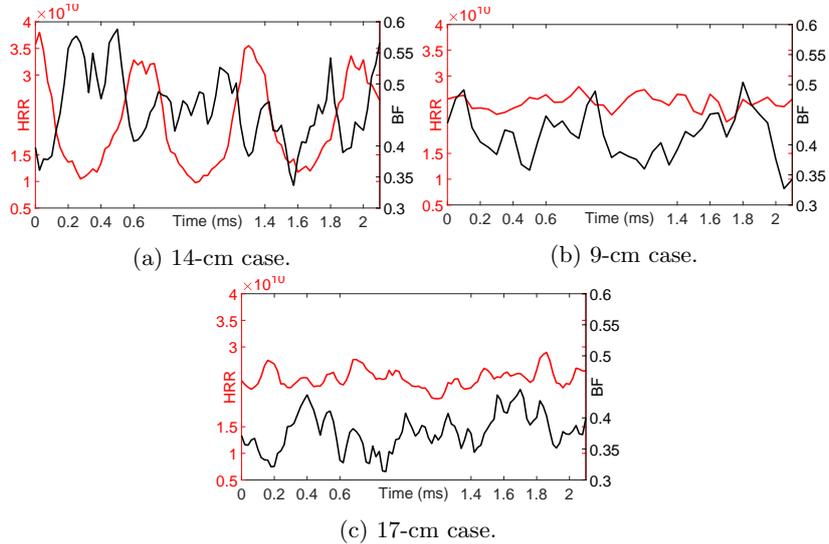

Figure 12: Volume-averaged HRR and non-premixed burning fraction (BF) for three different cases. HRR has the unit $J/m^3 s$.

The analysis above suggests that while the current CFPV approach can describe the partially premixed flame in the CVRC experiments, the premixed burning mode is dominant. This means a multiple regimes of non-premixed and premixed flamelet approach should be used instead of the current non-premixed flamelet based approach. However, there is also a weakness in the Takeno flame-index definition. By taking the dot products of only the global reactant gradients, it does not adequately represent the flame behaviors when a complex detailed mechanism is used. Following Seshadri and Peters [40], there exists a small diffusion control reaction layer around the stoichiometric mixture fraction on the fuel-rich side. As shown by Fiorina *et al.* [45] in the counterflow double flame configuration, the flame index cannot adequately distinguish this region from the adjacent premixed fuel-rich flame. This likely means the non-premixed fraction of the total HRR should also be higher.

## 5. Conclusions

Axisymmetric simulations of highly unsteady, nonlinear combustion dynamics inside a model liquid rocket engine have been performed. Turbulence is treated with the Delay Detached Eddy Simulation (DDES) model. Combustion is modeled using the extended compressible version of the Flamelet Progress Variable approach by Pierce and Moin [17]. The fuel is methane at standard conditions. Decomposed hydrogen peroxide at 1030 K is used as oxidizer. Steady laminar diffusion flamelet solutions are obtained using the FlameMaster program [38]. Flamelet behaviors under different conditions of the S-curve are examined. In the limit of higher pressure, the combustion process becomes more efficient, thus resulting in hotter flame. In the unstable flamelet burning branch, highest HRR is dominated by the non-equilibrium structure on the fuel-lean side. In the stable flamelet burning branch, the reaction zone shifts to the fuel-rich side.

In the unstable case (14-cm oxidizer post), a standing half-wavelength pressure wave is found in the combustion chamber. Pressure anti-nodes occur near the dump plane and the exit of the combustion chamber. A pressure node exists at mid-chamber. Using the Rayleigh index, strong coupling between the HRR and pressure is found at the upstream pressure anti-node. When the pressure peaks near the dump plane, an adverse-pressure gradient is imposed on the reactant streams. As a result, the flame moves upstream close to the injector lip. During this time, the flame is dominated by premixed burning. Significant local extinctions and reignitions occur during this period. As the pressure decreases inside the chamber, the flame moves further downstream and diffusion burning dominates, where less local ignition and extinction is found. An additional simulation with only stable flamelet burning branch was performed. Without local extinction and reignition, the flame anchored this case at the injector lip. Therefore, the pressure-HRR coupling significantly decreased compared to the simulation where the full S-curve was allowed.

Combustion dynamics are further examined for flames under different pressure instability conditions. In the semi-stable case (9-cm oxidizer post), the flame is lifted away from the injector lip and weakly anchored at the dump plane. In the stable case (17-cm oxidizer post), the flame moves further downstream. In both cases, there is no strong axial flame movement as previously found with the 14-cm oxidizer post. There is a monotonic decrease in local extinction and reignition as the flow becomes more stable (decreases in pressure fluctuations). However, extinction and reignition still occurs around the dump plane even for the stable burning. Therefore, the whole S-curve, and by extension the CFPV approach, should be utilized when flame/acoustic interactions are concerned.

Flame index analysis revealed the premixed flame as the dominant burning mode for all three cases. However, cautious interpretation of the flame index should be taken due to its oversimplified formulation. Nevertheless, there is still a significant amount of premixed burning. A hybrid premixed and non-premixed approach, like ones used by Knudsen and Pitsch [20], should be considered for future work. More accurate predictions of local extinction and reignitions should also be considered by using the Statistically Most Likely Distribution (SMLD) PDFs for either the progress variable or the pressure [18, 19].

**Acknowledgement**

This research was supported by the U.S. Air Force Office of Scientific Research under grant FA9550-15-1-0033, with Dr. Mitat Birkan as the program manager. Professor Heinz Pitsch of RWTH Aachen University is acknowledged for providing us access to the FlameMaster code.

**Appendix**

For clarification, a schematic of the solution procedure is also included. From know conditions at time n, the solution will be advanced to time n+1

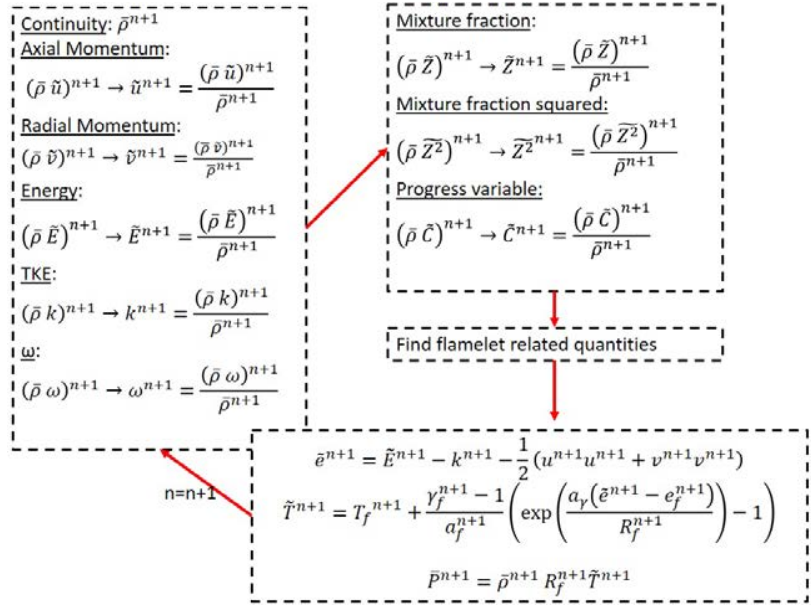

Figure 13

The code updates all of these conserved variables at the same time to ensure strong coupling.

The step of finding all the flamlet related quantities can be summarized as follows. Each quantity such as $(T_f^{n+1}, e_f^{n+1}, a_f^{n+1}, R_f^{n+1})$ has its corresponding pretabulated flamelet libraries (four-dimensional arrays). At each grid point in the computational domain, using the values of $\tilde{Z}^{n+1}, \widetilde{Z^2}^{n+1}, \tilde{C}^{n+1}, \bar{P}^n$, each quantity is computed from interpolation using its respective flamelet library.